# Microstrain induced deviation from Néel's 1/$d$ behaviour: Size-dependent magnetization in Bi$_{1-x}$Ca$_x$Fe$_{1-y}$Ti$_y$O$_{3-\delta}$ nanoparticles

**Pavana S.V. Mocherla[1], M.B. Sahana[2], Ehab Abdelhamid[3], Debarati Hajra[3], B. Nadgorny[3], R. Naik[3], R. Gopalan[2], M.S. Ramachandra Rao[4], B.R.K. Nanda[5] and C. Sudakar[1,*]**

[1] Multifunctional Materials Laboratory, Department of Physics, Indian Institute of Technology Madras, Chennai 600036, India.
[2] Centre for Automotive Energy Materials, International Advanced Research Centre for Powder Metallurgy and New Materials, IITM Research Park, Chennai-600113, India
[3] Department of Physics and Astronomy, Wayne State University, Detroit, MI 48201, USA
[4] Department of Physics, Nano Functional Materials Technology Centre and Materials Science Research Center, Indian Institute of Technology Madras, Chennai-600036, India
[5] Condensed Matter Theory & Computational Lab, Department of Physics, Indian Institute of Technology Madras, Chennai 600036, India.
*Corresponding author email ID: **csudakar@iitm.ac.in**

Magnetization of antiferromagnetic nanoparticles is known to generally scale up inversely to their diameter ($d$) according to Néel's model. Here we report a deviation from this conventional linear 1/$d$ dependence, altered significantly by the microstrain, in Ca and Ti substituted BiFeO$_3$ nanoparticles. Magnetic properties of microstrain-controlled Bi$_{1-x}$Ca$_x$Fe$_{1-y}$Ti$_y$O$_{3-\delta}$ ($y = 0$ and $x = y$) nanoparticles are analyzed as a function of their size ranging from 18 nm to 200 nm. A complex interdependence of doping concentration ($x$ or $y$), annealing temperature (T), microstrain ($\varepsilon$) and particle size ($d$) is established. X-ray diffraction studies reveal a linear variation of microstrain with inverse particle size, 1/$d$ nm$^{-1}$ (i.e. $\varepsilon d = 16.5$ nm.%). A rapid increase in the saturation magnetization below a critical size $d_c \sim 35$ nm, exhibiting a (1/$d$)$^\alpha$ ($\alpha \approx 2.6$) dependence, is attributed to the influence of microstrain. We propose an empirical formula M $\propto$ (1/$d$)$\varepsilon^\beta$ ($\beta \approx 1.6$) to highlight the contributions from both the size and microstrain towards the total magnetization in the doped systems. The magnetization observed in nanoparticles is thus, a result of competing magnetic contribution from the terminated spin cycloid on the surface and counteracting microstrain present at a given size. Large magnetodielectric response of ~ 9.5 % is observed in spark plasma sintered pellets with optimal size and doping concentration, revealing a strong correlation between magnetic and ferroelectric order parameters.

Magnetism in nanostructured $BiFeO_3$ offers a great deal of interest for many researchers working on multiferroics due to its intrinsic spin cycloid structure and elusive ferromagnetic signal. Dzyaloshinskii-Moriya (DM) interaction between the neighbouring $Fe^{3+}$ spins mediated *via* oxygen results in canting of antiferromagnetically (AFM) ordered spins. Weak net magnetic moments resulting from this uncompensated canting of neighbouring spin sub-lattices constitute the spin cycloid. This cycloidal structure propagates along $[10\bar{1}]$ direction with a repeat distance of 62±2 nm perpendicular to that of polarization direction along [111] in the material [1]. Any perturbation to this spin structure is believed to enhance the magnetization in an otherwise antiferromagnetic $BiFeO_3$, improving the possibility of better magnetoelectric coupling in the material. Several reports on nano $BiFeO_3$ confirmed an increase in magnetization with reduction of particle size. Park *et al.* [2] reported a spin-glass freezing behaviour due to interplay of finite size effects, inter-particle interactions, and a random distribution of anisotropy axes in $BiFeO_3$ nanoparticles of size ranging from 14 to 342 nm. Important observation of their studies also include the linear dependence of magnetization on the size of the nanoparticles (*d* nm), the well-known $1/d$ behaviour which can be explained using Néel's model [2-6]. Mazumder *et al.* [6] attributed the improved hysteresis with large coercivity, 450 Oe at 5 K, to the lattice strain-induced canting or ferromagnetism in nanometer (11 to 50 nm) sized $BiFeO_3$ particles. With reduction in size an increasingly diffuse Néel's transition accompanied by a shift in $T_N$ to lower temperatures has been reported widely and is commonly attributed to the combined effect of larger surface area, oxygen non-stoichiometry and strain in nanoparticle systems of $BiFeO_3$ [6-9]. Huang *et al.* [10] showed a deviation from $1/d$ behaviour at 65 nm, close to the repeat distance of spin cycloid in $BiFeO_3$, where a structural anomaly was observed along with an increase in magnetoelectric coupling. High value of magnetization, 0.4 $\mu_B$/Fe for 4 nm sized $BiFeO_3$ particles, was reported contrary to the bulk value of 0.02 $\mu_B$/Fe [4]. Jaiswal *et al.* [11]

attributed large splitting in field-cooled and zero-field-cooled M-T (magnetisation *vs.* temperature) curves to the frustrated spin structure and magnetic spin-strain interactions that mainly result from the excess strain, coordination distortion and lattice disorder in nanoparticles (55 nm). The reduction of size in $BiFeO_3$, thus, brings in intricately complex structural discrepancies due to the increased surface to volume ratio. It is important to note that the synthesis and calcination conditions play a crucial role in tuning the microstrain present in nanoparticles [12]. Under-coordinated oxygen defects and microstrain were found to be the two major sources of physical property variation at the nanoscale [9,12]. Strain engineering in materials, especially in ferroic oxides, is a novel approach to tune the physical properties [13-15]. Usually strain manipulation is explored in thin films where the strain related to the film-substrate interface is shown to control the magnetization [15,16]. An increase in magnetization with reducing thickness of $BiFeO_3$ films has been attributed to the influence of mismatch strain on the magnetic response. Similar studies were reported by Lim *et al.*[17] wherein a slight increase in magnetization (~ 5 emu/cm$^3$) of strain controlled $BiFeO_3$ thin films was observed. However, there are very few reports discussing the impact of microstrain on the physical properties of nanoparticle systems. Our previous studies on pure $BiFeO_3$ nanoparticles showed an abrupt drop in magnetization around a critical size ($d_c \approx$ 30 nm), accompanied by a shift in the magnetic Néel's transition [9]. This was shown to arise from the alteration of Fe-O-Fe chains across the nanoparticles due to the strain permeating from the surface. The magnetization trend in these pristine $BiFeO_3$ nanoparticles shows a clear deviation from 1/*d* behaviour at this critical size [9]. Thus, microstrain control is shown to be a novel approach for tuning physical properties of oxide materials.

In this study we present size-dependent magnetization of microstrain controlled $Bi_{1-x}Ca_xFe_{1-y}Ti_yO_{3-\delta}$ nanoparticle system. The microstrain is tuned by controlling the nanoparticle size and defects in $BiFeO_3$. Magnetization in these nanoparticles increases as their size decreases, more rapidly below ~ 35 nm exhibiting a $(1/d)^{\alpha}$ ($\alpha \approx 2.6$) dependence. This size-dependent magnetization deviates considerably from the Néel's $1/d$ linear variation [3]. Interestingly, microstrain ($\varepsilon$) shows a linear variation with $1/d$ ($\varepsilon d$ = 16.5 nm.%) between ~ 0.1 to 1.1 %. This microstrain is shown to influence the magnetization according to the empirical relation M $\propto (1/d)\varepsilon^{\beta}$ ($\beta$ = 1.6), which highlights the contributions from both the size and microstrain towards the total magnetization.

**Experimental**

Pure BiFeO$_3$ nanoparticles ($x = y = 0$) of sizes ranging from 22 nm to 65 nm were synthesized *via* low temperature citrate sol-gel method by controlling the annealing temperature [18]. Bi(NO$_3$)$_3$.5H$_2$O and Fe(NO$_3$)$_3$.9H$_2$O solutions of 0.2 M each were prepared in de-ionized water. Dilute nitric acid was used for better solubility of the precursors. These two solutions were mixed and citric acid is added to this mixture in 1:1 ratio to metal cations. This pale-yellow coloured solution was stirred constantly at 80 °C to 100 °C till a deep-brown gel is formed. The as-obtained precursor was calcined at different temperatures to control the size (referred to as BFO-$d$, $d$ is the average crystallite size in nm) [12]. Bi$_{1-x}$Ca$_x$Fe$_{1-y}$Ti$_y$O$_{3-\delta}$ nanoparticles were synthesized using the same method. CaCl$_2$.2H$_2$O and Ti-isopropoxide (Ti[OCH(CH$_3$)$_2$]$_4$) were used for Ca and Ca-Ti doping in BiFeO$_3$ in equimolar proportions with Bi and Fe precursors. The dried precursors obtained from the above synthesis were calcined under controlled heating conditions at temperatures T = 550 °C, 650 °C and 800 °C, and are referred to as BCFO-T/BCFTO-T [19].

The x-ray diffraction data on pure BiFeO$_3$ were taken using PANalytical X'pert PRO diffraction system and doped BiFeO$_3$ using Rigaku diffractometer with Cu-K$\alpha$ ($\lambda$=1.5406 Å). Crystallite size was calculated using Scherrer's formula after correcting for the instrumental broadening. The microstrain values are estimated from the single peak analysis using a pseudo-Voigt function in X'Pert highscore which fits Gaussian and Lorentzian components separately [20]. Raman spectra were acquired at room temperature with a Horiba Jobin-Yvon (HR 800UV) micro-Raman using a 632 nm excitation line from a He-Ne laser. Transmission electron microscopy (TEM) images, high resolution transmission electron microscopy (HRTEM) images and selected area diffraction (SAD) patterns are carried out using a Phillips CM12 operating at 120 kV and TECNAI T20 electron microscope operating at an applied voltage of 200 V. Magnetic hysteresis (M *vs*. H) curves were measured at room temperature (300 K) using EZ9 Microsense Inc. USA vibrating sample magnetometer up to 2 T and at low temperature (20 K) up to a maximum field of ±7 T using a Quantum Design SQUID vibrating sample magnetometer (SQUID VSM). Zero-field cooled and field-cooled measurements were also done on SQUID VSM. The ferroelectric PE loops and leakage current characteristics on these pellets were carried out using a Precision Premier-II ferroelectric loop tracer by Radiant technologies. The in-built 100 V voltage source with an additional TReK model 609E-6 high voltage amplifier provides a maximum voltage limit of 4 kV. Frequency and temperature dependent dielectric studies in a parallel-plate capacitor configuration were carried out on a dielectric spectrometer of Novocontrol Technologies. Similarly, the capacitance and dielectric loss tangent (tan δ) of the silvered pellets were acquired using Agilent 4284A LCR meter at 1 MHz frequency and $V_{peak}$ = 1 V. The magnetodielectric behaviour was measured with a maximum applied field of 5 T controlled by Quantum Design Physical Property Measurement System (PPMS) at 10 K. Initially, the sample is loaded in the PPMS at room temperature and cooled to 10 K under zero magnetic field. Once the temperature is stabilized, the magnetic field was swept twice from +5 Tesla to -5 Tesla at a rate of 50 Oe/sec. The magnetic field, sample capacitance and dielectric loss were recorded using LabVIEW program.

**Results and discussion**

$Bi_{1-x}Ca_xFe_{1-y}Ti_yO_{3-\delta}$ commonly represents two systems: (i) $Ca^{2+}$ is substituted at $Bi^{3+}$ site, $Bi_{1-x}Ca_xFeO_{3-\delta-x/2}$ (BCFO: $x \neq 0$, $y = 0$; $x/2$ is the oxygen vacancy concentration produced due to divalent cation substitution) and (ii) $Ca^{2+}$ is substituted at $Bi^{3+}$ site and $Ti^{4+}$ is substituted at $Fe^{3+}$ site in equi-molar proportions, $Bi_{1-x}Ca_xFe_{1-y}Ti_yO_{3-\delta}$ (BCFTO: $x = y$). In both the cases, $\delta$ represents the small intrinsic oxygen defect concentration present in $BiFeO_3$ nanoparticles irrespective of doping.

It was found that the doping concentration, calcination temperature, microstrain and average particle size exhibit a systematic correlation with each other [19]. Fig.1a and 1b show the particle size as a function of doping concentration and microstrain variation with particle size, respectively. The average particle size is seen to decrease monotonically with increasing doping concentration (Fig.1a). This change is much more rapid in BCFO samples than in BCFTO. The difference between the size variations in these two systems is detailed further in the following sections. A striking feature obtained from the structural analysis is the observed linear variation of microstrain with inverse particle size ($1/d$) (i.e., $\varepsilon d \approx 16.5$ %.nm) (inset of Fig.1b). Microstrain also shows a linear change with doping concentration ($x$ or $y$), however the slope of this variation is higher in BCFO than in BCFTO [19]. Raman spectral studies from our previous work on $BiFeO_3$ nanoparticles revealed systematic changes in the peak positions and relative intensities of characteristic two-phonon modes which are correlated to the microstrain present in the system [9]. This band, which is an overtone of fundamental Fe-O vibrations at low frequencies, is very sensitive to any disturbance in the alignment of Fe-O-Fe bonds. The two-phonon mode in both BCFO and BCFTO is de-convoluted and the peak positions of selected samples are analyzed as a function of size (Fig. 1c). The peak at ~ 940 cm$^{-1}$ gradually shifts to higher frequencies with reduction in particle size, marked by a rapid shift at ~35 nm (Fig. 1d). This variation is in close resemblance to the dependence of microstrain on size (Fig. 1b). Thus, it is evident that the microstrain present in the nanoparticles below a critical size ($d_c \approx 35$ nm), influences the Fe-O bond characteristics. However, it should be noted that peak positions in samples annealed at 650 °C and 800 °C suffer a slight deviation to the observed trend. This could be due to the change in the structure in addition to contributions of surface and core defects to the total microstrain with increase in size. In any case, separating the influence of individual components of microstrain on the Raman modes is not possible. All these observations indicate a strong dependence of microstrain on size, with an additional contribution from the defects, if present. Thus, strain in nanoparticles is an aggregate of two sources: (i) surface strain due to reduction in particle size and (ii) localized strain originating from a large number of oxygen vacancies in the core (refer to Fig. 2 for details).

There are two major factors which influence the particle size. One is the crystallization temperature and the other is the oxygen vacancy concentration. We observed from thermal studies that the crystallization of doped samples happens at much higher temperature than pure BFO (Fig. 3). Hence, BCFO samples are of smaller size than BFO under identical annealing conditions. In addition to this fact, $Ti^{4+}$ doping in BCFTO resulted in smaller particle sizes than BFO and BCFO. It has been reported earlier that $Ti^{4+}$ doping in BFO reduces the particle size far less than the pure BFO [21,22]. Reetu *et al*. [23] reported that $Ti^{4+}$ doping at $Fe^{3+}$ site in BCFO suppresses the formation of oxygen vacancies. This in turn slows down the oxygen ion motion and reduces the grain growth. This results in the observed size difference between BCFO and BCFTO with the latter being smaller. In the present study, the size range of BCFTO is almost half of that in BCFO for the same doping and calcination conditions, which indicates that the surface area is more in BCFTO than in BCFO. Hence, for a given doping concentration the microstrain in BCFO system should be much smaller than that of BCFTO. Instead we observed very little difference in their corresponding microstrain values. This suggests that BCFO samples have additional perturbations appending the surface changes. We show that the oxygen vacancies present in BCFO, due to charge imbalance, are giving rise to these core perturbations, whereas the local strains produced by $Ti^{4+}$ ($r_{ionic}$ = 60.5 pm) substitution at $Fe^{3+}$ ($r_{ionic}$ = 64.5 pm) site is insignificant due to a small difference in their ionic size values. At 550 $^o$C, these vacancies are believed to be randomly distributed due to non-homogeneous substitution of $Ca^{2+}$ at low temperatures. However, transmission electron microscopy (TEM) and selected area electron diffraction (SAED) studies confirm the ordering of these oxygen vacancies within the nanocrystallites at higher temperatures, which is attributed to the strain relaxation [19].

Contrast variations within the crystallites due to localized strain fields are evidenced from the bright field transmission electron microscopy images (Fig. 4). A clear change in the average crystallite size of BCFO and BCFTO compared to that of pure BFO is also observed. High resolution transmission electron microscopy (HRTEM) images of pure BFO ($d$ = 22 nm), BCFO-650 ($x$ = 0.05; $d$ = 57 nm) and BCFTO-650 ($x$ = $y$= 0.05; $d$ = 32 nm) are shown in Fig. 5. Filtered inverse fast Fourier transformed (FFT) images of these representative crystallites clearly show the differences in the nature of strain variation among these samples. While pure BFO samples show the existence of strained planar arrangements near the surface, BCFTO shows strained regions due to surface modifications permeating into the core of the crystallites as well. BCFO on the other hand reveals a larger distribution of core strains resulting due to the presence of oxygen vacancies. The strain originating from these vacancies relaxes *via* vacancy ordering for samples annealed at higher temperatures (> 600 $^o$C) [19]. These observations imply that microstrain in BCFO has contributions both from surface and core defects while it is predominantly a surface property in BCFTO. The contrast variations within these doped systems, unlike in the pure BFO nanoparticles, is thus attributed to the existing strain fields. As the annealing temperature increases, particle size increases relaxing the microstrain in nanoparticles.

BCFO and BCFTO samples showed that Fe is in 3+ oxidation state for all the doping concentrations, as discerned from XANES studies [19], indicating that the charge deficit in $Ca^{2+}$ substituted samples is compensated by oxygen vacancies. Changes in the high energy region of O-K edge reflect the modifications in the oxygen ion neighborhood due to doping. Besides, Ti edge in BCFTO shows features similar to that of an octahedral environment confirming the $Ti^{4+}$ substitution at $Fe^{3+}$ sites. The peak positions in both the Fe-L and O-K edges of BCFO revealed fine modifications in the Fe-O-Fe octahedral arrangements in presence of $Ca^{2+}$, while the close matching of these values in BCFTO with that of BFO affirmed the near-stoichiometric and relatively oxygen-vacancy-free environment [19].

Fig. 6 shows the room temperature magnetization *vs.* magnetic field (M-H) plots for representative samples of BCFO and BCFTO annealed at 550 °C, 650 °C and 800 °C. The saturation magnetization (ferromagnetic component, $M_f$), obtained after subtracting the linear increase in M above 2 T, is shown in Fig. 7 for varying *x* and *y* concentrations in $Bi_{1-x}Ca_xFe_{1-y}Ti_yO_{3-\delta}$. M-H curves of BCFO-550 and BCFTO-550 (Fig. 6) show saturated behavior indicating the presence of large ferromagnetic component in these samples. This could be due to relatively smaller crystallite size resulting from doping at low temperatures (~550 °C) compared to the pure $BiFeO_3$ annealed under same conditions. A large fraction of surface atoms in these nanoparticles (< 60 nm) in addition to the core uncompensated spins due to doping and subsequent local termination of spin cycloid could be the sources for increased magnetization (refer to Fig. 9). Reports on Ca doped $BiFeO_3$ annealed at low temperatures (< 600 °C) also showed an increase in magnetization due to enhanced canting resulting from the buckling of Fe-O-Fe bond angle [24]. However, a comparison of magnetization in BCFO with that of BCFTO shows that oxygen vacancies and surface strain counteract the increase in magnetization of BCFO.

On the other hand, BCFO-650 and BCFO-800 show decreased magnetization about one and two orders of magnitude respectively, compared to BCFO-550. Nevertheless, the increase in magnetization with '$x$' is retained in all these samples. The M-H plots show a non-saturating behavior even at 7 T indicating a large fraction of antiferromagnetic component overlapping the weak ferromagnetic contribution. The magnetization in Ca doped BFO samples as reported by Bhushan *et al*. [25] (nanoparticles of size 28 nm) is 2.38 emu/g, by Khomchenko *et al*. [26] (polycrystalline bulk) is 0.05 emu/g, by Feng *et al*. [27] (nanofibers) is ~ 0.6 emu/g. The low temperature magnetization in our BCFO-550 (20 nm) sample is ~ 2.0 emu/g and BCFTO-550 (18 nm) is ~ 4.1 emu/g. The values are close to that reported by Bhushan *et al*. [25] for nanoparticles and much higher than previously reported bulk values (Fig. 7). Thus, despite the apparent increase in magnetization with doping concentration *$x$*, it is understood that the actual influence is due to the change of particle size that results from doping. Magnetization in BCFO-800 is found to be the lowest of all the samples confirming the influence of size and reduced surface area on annealing at higher temperature.

The magnetization values at low temperature (20 K) and room temperature (300 K) are of the same order, ruling out the presence of any ferromagnetic impurities like $Fe_2O_3$ and $Fe_3O_4$. Even if these impure phases are present in traces a minimum magnetization of ~ 10 to 15 emu/g would have resulted [28-30], which is not the case in any of the samples under current study. This is also supported by the absence of any magnetic impurity phase related modes in Raman spectra in both BCFO and BCFTO (spectra not shown) samples and further substantiated by the 3+ oxidation state of Fe in these samples as evidenced from the XPS and XANES studies [12,19]. Hence any increase in magnetization of doped sample is considered to be an intrinsic variation of particle size and related surface effects and not due to any ferromagnetic secondary phases. The saturation magnetization values of these samples ($M_f$) increase steadily with doping concentration '$x$ ($=y$)' for both BCFO and BCFTO (Fig. 7a and 7b). It is important to note that the magnetization of BCFTO is an order of magnitude higher than their BCFO counterparts. BCFTO-550 nanoparticles with $d \sim 30$ nm show six times higher magnetization ($\sim 3.0$ emu/g) than BCFO-550 ($\sim 0.5$ emu/g) samples with similar size scale. This reveals that microstrain caused by oxygen vacancies in BCFO samples has a larger impact on the magnetic order than the surface strain produced by mere size reduction in BCFTO.

Magnetization *vs*. temperature (M-T) curves measured under zero-field-cooled (ZFC) and field-cooled (FC) conditions with an applied magnetic field of 500 Oe are shown in Fig. 8, for $x = 0, 0.05, 0.1$ ($y = 0$) and $x = y = 0, 0.05, 0.1$. Field-cooled curves show flat temperature dependence in all samples in addition to a significant splitting between ZFC and FC curves (Fig. 8). Absence of any peaks in these curves rules out the existence of magnetic impurities or superparamagnetic features in pure and doped BFO nanoparticles. Samples annealed at high temperatures show reduced magnetic moments compared to those annealed at 550 °C. Kinks observed around 120 K and 250 K of 800 °C annealed samples could be due to the Bi loss at high temperatures. Especially, the kink observed at 250 K is the magnetic Néel's transition of the Bi-deficient secondary phase $Bi_2Fe_4O_9$. These features do not contribute to the overall magnetization as this phase is originally antiferromagnetic and becomes paramagnetic after 250 K. The splitting between the ZFC and FC curves increases with a decrease in particle size. The M *vs*. T plots reflect the temperature dependence of weak ferromagnetism arising due to the uncompensated moments from cycloidal spin structure. However, these ZFC/FC magnetization values would also comprise a small fraction of paramagnetic component arising from disorder in the systems which show an upturn in the magnetization plots (please refer to the BFO-800 and BCFO-800 data in Fig. 8a & 8b). Such disorder could kill the AFM interaction of Fe-O-Fe resulting in a small fraction of paramagnetic components on the surface. This anomaly is seen pronounced in high temperature annealed samples (large sized samples), where the contribution from uncompensated moments becomes small with increased fraction of AFM phase. This upturn is not observed in the smaller particles because the magnetization due to the uncompensated surface moment is dominant.

The oxygen vacancies in BCFO suppress the spin cycloid propagation causing local magnetic disorder (Fig. 9). This could be the reason for higher magnetization observed in BCFO than pure BFO. However, oxygen vacancy-free, size-controlled BCFTO nanoparticles show larger magnetization than BFO and BCFO. This is attributed to the large surface ferromagnetism and less magnetic disorder in these nanoparticles compared to BCFO (Fig. 9). The presence of Ti at Fe site breaks the long range spin cycloid without structural distortion and leaves large uncompensated moments arising from Fe($d^5$)-O-Ti($d^0$) spin arrangements. Thus, magnetization in $Bi_{1-x}Ca_xFe_{1-x}Ti_xO_3$ nanoparticles arises from two main contributions: (i) the surface terminated spin cycloid structure (due to size reduction) and (ii) the local magnetic disorder produced in the core either due to oxygen vacancies (in BCFO) or $Ti^{4+}$ substitution at $Fe^{3+}$ site (BCFTO). The particle size of BCFO is larger (almost twice) than that of BCFTO, i.e., the surface area of BCFTO is four times larger than that of BCFO. Hence the magnetic contribution due to (i) is smaller in BCFO. In BCFTO, contribution (i) dominates due to their small size. In both cases, the contribution from (ii) is present. Hence BCFTO exhibits higher magnetization than BCFO. ZFC-FC measurements established that the magnetization observed in all the samples is a result of suppressed spin cycloid and related size effects.

It is reported that Fe-O-Fe bond angle (154.9° for bulk $BiFeO_3$) has a strong influence on its magnetic properties. Recent calculations by Modak *et al.*[31] examined the magnetic configuration on a small cluster of size ~ 1.1 nm and suggested that there could be a magnetic phase transition from antiferromagnetic to ferromagnetic state for a Fe-O-Fe bond angle of 133°. However, such a wide deviation in bond angle is not realized experimentally as it is concurrent with a structural transition in nanoparticle system [32-35]. Experimental studies that report Fe-O-Fe bond angle show a variation by 2 to 3° and a net change in magnetization around few tens of milli emu/g which is insignificant [4,36-39]. As per the Rietveld refinement done on our samples, Fe-O-Fe bond angle changes by a maximum of 4° from the bulk value. Thus, we believe that crystal lattice distortion due to the Fe-O-Fe bond angle variation that might be present in doped systems has a negligible effect on the total magnetization. In other words, ignoring the size related surface effects as in the case of bulk samples (samples annealed at 800 °C) and considering only the effect of Fe-O-Fe bond angle will result in a minimal change of the net magnetization. Therefore we attribute the observed magnetization in the present study to the predominant size related surface effects.

Out of the four correlated variables ($x/y$, T, $\varepsilon$, and $d$) discussed in the previous sections, particle size ($d$) and microstrain ($\varepsilon$) have a direct correlation with each other (Fig. 1b), hence we presume that the attempt of analyzing microstrain as a function of particle size seems to be a fair choice. From the microstructural and magnetic property studies discussed so far, we also infer a strong correlation of magnetic order with the particle size $d$. $M_f$ is plotted as a function of particle size irrespective of dopant concentrations ($x$ or $y$) in Fig. 10a. This plot unambiguously suggests that the magnetization is a function of size irrespective of different combinations of doping and calcination temperatures. A sharp increase in magnetization below ~35 nm in our study is significantly different from the increase in magnetization seen in nanostructures reported earlier [2,4,6]. In general, the surface to volume ratio in spherical nanoparticle of radius $r$ is $4\pi r^2 a/(4/3)\pi r^3 = 3a/r$, where $a$ is the interatomic spacing and $r$ is the radius of the particle [40]. Thus, the fraction of surface atoms in antiferromagnetic nanoparticles goes by $1/r$ or $1/d$ [40]. The magnetization of a spherical nanoparticle BiFeO$_3$ of radius $r$ as suggested by Zhang *et al*. [41] is given as

$$M = 3M_o \sin\theta_o \frac{(qr\cos qr - \sin qr)}{(qr)^3} \quad \ldots\ldots\ldots\ldots (1)$$

where, $M_o$ is the maximum unit cell magnetization, '$q$' is the spin cycloid wave vector, $\theta_o$ is the initial phase angle, and '$r (= d/2)$' is the radius of the nanoparticle. This equation holds well for a sample with ideal particles of a definite size '$d$' without any surface disorder and defects. However, experimental synthesis conditions invariably produce nanoparticles with a range of size, usually defined by a lognormal distribution

$$P_d(x) = \frac{A}{\sqrt{2\pi}Sx} \exp\frac{-[\ln(x/d)]^2}{2S^2} \quad \ldots\ldots\ldots\ldots (2)$$

where $S$ is the standard deviation, $d$ is the mean particle size and A is the area under the distribution curve. Therefore, the net theoretical macroscopic magnetization $M_{theory}(d)$ is

$$M_{theory}(d) = \sum_{x=1}^{n} P_d(x) M(x) \quad \ldots\ldots\ldots\ldots (3)$$

Here $M(x)$ is the magnetization of a BiFeO$_3$ particle of specific size $x$ as given by equation (1). Most of our samples exhibit a well-defined lognormal distribution for particle size with a standard deviation $S = 0.3$. The theoretically calculated magnetization from equation (3) is plotted in Fig. 10a for comparison with the experimental data. It is worth noting that for different values of S only the shape of the magnetization curve changes and the magnetization for $d < 50$ nm remains largely unaffected (Fig.11). This magnetization decreases smoothly with increase in particle size unlike the oscillatory behaviour reported beyond 80 nm by Zhang *et. al*. [41]. While theoretically calculated magnetization agrees well with the typical $1/d$ variation of magnetization presented in experimental results within a certain range of size 20 to 50 nm [2,6], our results significantly deviate from this linear fit (Fig. 10b inset). Rather, the magnetization in our case changes much slower initially, until it starts to increase rapidly at ~ 35 nm. This reduction in size influences the magnetization to a considerable extent.

The Néel's model which describes the nanoparticles (of antiferromagnetic bulk) as having a ferromagnetic shell covering the anti-ferromagnetic core fits suitably with the 1/$d$ dependence of magnetization on size [3]. In fact, several reports presented a linear variation of magnetization in BiFeO$_3$ nanoparticles with 1/$d$ [2,6,11]. However, there are also examples of few deviations for certain particle sizes which were attributed to either strain [9] or structural anomalies specific to that size [10]. It was observed from our earlier studies that pure BFO nanoparticles of size around 30 nm showed an abrupt fall in magnetization due to the microstrain [9]. It has been very evident that the increased surface area below a certain size is large enough to perturb the core stoichiometry. It was understood that below a critical size, the magnetic structure can no longer be explained by Néel's model since the strain that develops on the surface gradually propagates into the core of the particle altering the bonding and coordination [2,42]. Nevertheless, a systematic study of microstrain in BiFeO$_3$ nanoparticles has not been performed and its subtle influence on the physical properties is less noticed.

In the current situation, we observe a significant deviation from this linearity in $M_f$ vs. 1/$d$ (inset of Fig. 10b). In fact, log $M_f$ vs. log(1/$d$) could be fitted to a linear function for our samples (Fig. 10b). Based on these inputs we arrived at an empirical formula, $M_f \propto (1/d)^{\alpha}$, where $\alpha = 2.6$ is the slope obtained from log $M_f$ vs. log (1/$d$) plot. Since microstrain ε varies linearly as a function of inverse particle size 1/$d$ (i.e., ε$d$ = 16.5 %.nm), the above relation can be re-written as M $\propto$ (1/$d$).(1/$d$)$^\beta$ (where β = α-1), i.e., $M_f \propto (1/d).(\varepsilon/16.5)^\beta$ or **$M_f \propto (1/d).\varepsilon^\beta$**. As $d \rightarrow \infty$ and ε $\rightarrow$ 0, i.e., for large particles where the strain is negligible the ferromagnetic contribution is almost zero. The other extreme value of $M_f$ becoming infinitely large is a non-physical condition since the magnetization is limited to have definite values of '$d$' and 'ε'. It is well-known that microstrain depends strongly on the synthesis parameters. Hence, if it is possible to prepare the nanoparticles of size smaller than 62 nm with minimal microstrain, then the commonly observed Néel's $1/d$ behaviour can be restored with the exponential contribution of microstrain becoming insignificant ($\alpha \rightarrow 1$).

The relation $M_f \propto (1/d)\,\varepsilon^\beta$ with $\beta \approx 1.6$ is verified with the microstrain values obtained from x-ray diffraction (XRD) and is seen to fit very well with our magnetization data as shown in Fig. 10a. This unequivocally establishes the influence of microstrain on the observed magnetization for the given particle size range. However, it should be noted that the value of $\beta$ is specific to the dependence of microstrain and magnetization on particle size. There are several reports which discuss the influence of strain on the structure and magnetic nature of materials [43-45]. Most of the studies relate to the lattice strain observed in thin films in this regard. Nevertheless, the fact that antiferromagnetic materials show an increased magnetic moment due to the strain is well-established. Antiferromagnetic $LuMnO_3$ shows a moment of $\sim 1\mu_B$ at the strained substrate-film interface [46]. Lim *et al.* [17] also showed strain enhanced magnetization in $BiFeO_3$ thin films. In the present study, as the microstrain itself is strongly dependent on particle size, an extrapolation of these observations supports the point that microstrain caused by defects in the lattice has a dominant effect *only* when the particle size is adequately small. The surface strain permeates well into the nanoparticle core below a critical size $d_c$ (~35 nm) and the magnetic order below this value experiences a conflated effect of the microstrain induced from both the surface and localized oxygen vacancies. Large magnetization in BCFTO indicates that these systems have substantially reduced microstrain compared to BCFO even though they are annealed at same temperature. Again, we emphasize that the absence of oxygen vacancies and increased contribution from the surface and core uncompensated spins are the reasons for the observed large magnetization in BCFTO. This is a better demonstration of spin cycloid termination below 62 nm compared to the BCFO.

Owing to the intrinsic coupling between the magnetic and ferroelectric order parameters, any change in the magnetization would invariably result in altered ferroelectric characteristics and magnetoelectric coupling in $BiFeO_3$. So, we worked further on verifying the size and strain dependence of ferroelectric properties of BCFO/BCFTO nanoparticle systems, having seen a systematic influence of size and strain on the magnetization. However, studying the size-dependent ferroelectricity of nanoparticle systems is extremely challenging and not straightforward. In order to measure the polarization, it is required to compact the powder samples into dense pellets. This is conventionally done by sintering at high temperatures wherein a rapid growth of particles takes place. Measurements on such pellets do not produce the nanoscale ferroelectric properties. Therefore, we have adopted spark plasma sintering (SPS) technique by which highly dense pellets with nano sized crystallites can be obtained. The as-prepared samples of pure and doped $BiFeO_3$ (only 5 % and 10 %) were annealed at 550 °C initially for 2 h and plasma sintered at 650 °C for 5 min to achieve highly dense pellets. The density of these pellets as estimated from the displacement of plungers is found to be ~ 96 to 99 %. The as-obtained pellets were re-annealed at 550 °C in air for 15 min to remove any residual carbon from the graphitic die. SEM studies have confirmed that the re-annealing process has not affected the grain size. The SEM images (Fig. 12) taken on these pellets reveal that the grain size of pure $BiFeO_3$ increases rapidly (400 nm to 2 μm) from that of the powder value (60 to 70 nm). For BCFO and BCFTO samples (5 % and 10 %), the size still remains in the nanoregime. Grain size in BCFO05 ($x = 0.05, y = 0$) is found to be between 100 to 170 nm and BCFO10 ($x = 0.1, y = 0$) show much smaller size between 50 to 90 nm. On the other hand, BC05FT05 ($x = y = 0.05$) has size ranging from 35 to 65 nm and BC10FT10 ($x = y = 0.1$) varies between 40 to 90 nm.

It is difficult to estimate the exact crystallite size from SEM studies; the average crystallite size and microstrain values estimated from XRD studies are given in Table 1. This large difference in the values determined from XRD and SEM is due to agglomeration of nanocrystallites under high temperature and high pressure conditions provided during the plasma sintering process. Microstrain in its usual trend is found to be larger for smaller crystallite sizes. However, unlike in powders, a fine control of size and strain is not possible as these values are subject to modification during sintering.

While pure BFO and BCFO show typical hysteresis of ferroelectric material with a maximum polarization of 0.2 μC/cm$^2$, BCFTO samples show very leaky behaviour even for modest voltages ~ 200 to 300 V (Fig. 13a). The round shape of the loops reflects the leaky nature of the sample. Leakage current characteristics (J *vs*. E) in Fig. 13b show that BCFO samples have lesser leakage than BCFTO. Despite the lack of oxygen vacancies in BCFTO, the small grain size and large number of boundaries could be responsible for conduction in these samples. These current density data are fitted for straight lines. Low voltage curves show good fits with slopes close to 1 indicating ohmic type conduction. High voltage curves give a slope ~ 2 indicating a trap-filled-limited conduction [47,48]. Temperature and frequency dependent dielectric variation was carried out on these SPS pellets (figures not shown). The low temperature dielectric constant values at 100 K at 1 MHz frequency for BFO, BCFO05, BCFO10, BC05FT05, and BC10FT10 samples are ~ 60, 192, 150, 131, and 205 respectively. The large dielectric constant observed in BC10FT10 is attributed to the nanosized grains and large number of boundaries in the sintered pellet [49].

All the samples show a smooth variation of dielectric constant with respect to the applied magnetic field (Fig. 14). Percentage change in magnetodielectric response is given by

$$MD \% = \left[\left(\frac{\varepsilon(H)}{\varepsilon(0)}\right) - 1\right] \times 100 \% \dots\dots\dots\dots(4)$$

where ε(H) is the dielectric constant at field H and ε(0) is the dielectric constant under zero field. It is seen that both BCFO05 and BC05FT05 samples show a very large change in MD response ~ 9.5 % compared to pure BFO which shows only 0.06 %. Surprisingly, BCFO10 and BC10FT10 samples show very small MD change of ~ 0.05 and 0.9 % respectively. The possible contribution of magnetoresistive and interface effects to the overall magnetodielectric response as suggested by Catalan [50] can be ruled out in the current study since all the measurements were done at 1 MHz. In an ideal situation the MD response should increase when the particle reduces to a size smaller than the spin cycloid length (62 nm) due to the increasing ferromagnetic component. However, we see that there is no systematic correlation of the MD response to the size of the sample contrary to the previous reports [49]. Another considerable factor which plays an important role in altering the MD value is the structural distortion induced by doping. In the present study, the MD response is found to decrease with increase in doping concentration. This trend suggests that there is an optimum range for doping concentration and size in which the magnetodielectric response attains a maximum value. For BCFO05 (18 nm) and BC05FT05 (26 nm) samples, the size is small enough to produce weak ferromagnetism and improve the coupling with the polarization whereas for BCFO10 (14 nm) and BC10FT10 (15 nm) the size reduces to smaller values where the ferroelectric aspect seems to be lost. Selbach *et al.* [51] have estimated a critical size of ~ 10 nm down to which the ferroelectricity can be retained in rhombohedral $BiFeO_3$ nanoparticles. However, many significant parameters in these sintered pellets like grain size uniformity, strain and other interface effects at the boundaries, etc. alter the intrinsic ferroelectric property far beyond than understood. Hence, a direct comparison of the SPS pellets to the nanoparticle powder samples cannot be made. The variation of MD response with respect to $M^2$ has also been plotted (Fig.15). This plot shows the linear variation of MD response above 5000 Oe. Lawes *et al.* [52,53] have reported that the scalar biquadratic term $P^2M^2$ in the free energy expansion $F = 1/2\varepsilon_o P^2 - PE - \alpha PM + \beta PM^2 + \gamma P^2M^2$ cannot be valid in antiferromagnetic materials with limited magnetic ordering and indicated the possibilities of observing high magnetodielectric response even in the absence of magnetoelectric coupling. In the current study, although we believe that the observed response is from the BCFO/BCFTO system, a detailed analysis of dielectric behaviour around the magnetic transition temperature (~ 643 K) is required to confirm the magnetoelectric coupling in these sintered pellets. In summary, ferroelectric characteristics and the magnetoelectric coupling could not be studied as a function of size and microstrain like in the case of powder nanoparticle systems. However, we confirm that by carefully altering the sintering conditions, and choosing an optimum combination of doping concentration and size can yield good magnetodielectric reponse in SPS pellets.

**Conclusion**

A detailed analysis of microstrain in BCFO and BCFTO nanoparticles in comparison with the pristine BFO nanoparticles is carried out over a wide range of doping concentrations annealed at different temperatures. Particle size and microstrain can be controlled by tuning the synthesis and annealing conditions. A linear dependence between the estimated microstrain and inverse particle size for different dopant and temperature combinations is revealed. The high resolution transmission electron microscopy images have shown lattice distortions due to local strain within the crystallites. Magnetization found in our doped systems is much higher than the reported bulk values indicating the remarkable influence of size controlled *via* doping. While the spin cycloid termination below 62 nm tends to produce a weak ferromagnetism on the nanoparticle surface, the microstrain developed due to surface modification and defects counteracts to randomize the spins and infuses into the core of nanoparticles below a critical particle size ~ 35 nm. This results in the deviation of magnetization away from the expected $1/d$ variation which is given by an empirical relation $M_f \propto (1/d)\varepsilon^{\beta}$ with $\beta \approx 1.6$ that counter-relates the microstrain and magnetization. The dielectric measurement in presence of magnetic field on spark plasma sintered samples show unusually large magnetodielectric response of 9.5 % in Ca and Ti doped (5 *at*. %) $BiFeO_3$, thus indicating a strong coupling between the magnetic and ferroelectric order parameters in nanosized samples.

**Acknowledgments**

PSVM and CS acknowledge the Central Microscopy facility and SQUID VSM facility at IIT Madras. PSVM acknowledges UGC for the financial aid. The work at Wayne State is supported by the NSF DMR grant # 1306449.

**Author contribution statement:** CS and MSR proposed the project. CS and PSVM designed and conducted the experiments. MBS and RG helped in room temperature magnetic measurements. CS proposed the empirical formula. The theoretical component is performed by BRKN. PSVM and CS analyzed the data and wrote the paper. Magnetodielectric measurements were carried out at Wayne State University, US. All authors discussed and revised the final manuscript.

**Competing Financial interests:** The authors declare no competing financial interests.

**Table 1:** Average crystallite size and corresponding microstrain values estimated from the x-ray diffraction patterns of SPS pellets. The crystallite size values are found to be much smaller than the grain size observed in SEM images.

| Sample | Grain size from SEM (nm) | Avg. crystallite size from XRD $d$ (nm) | Strain $\varepsilon$ (%) |
|---|---|---|---|
| Pure BFO ($x = y = 0$) | > 400 | 50 | 0.36 |
| BCFO05 ($x = 0.05, y = 0$) | 100-170 | 23 | 0.6 |
| BCFO10 ($x = 0.1, y = 0$) | 50-90 | 18 | 0.92 |
| BC05FT05 ($x = 0.05, y = 0.05$) | 35-65 | 21 | 0.67 |
| BC10FT10 ($x = 0.1, y = 0.1$) | 40-90 | 14 | 1.05 |

**FIGURE CAPTIONS**

**Fig. 1 (Color online):** (a) Variation of crystallite size ($d$ nm) in $Bi_{1-x}Ca_xFe_{1-y}Ti_yO_{3-\delta}$ as a function of $x$ (=$y$) $at.$ % (b) Microstrain ($\varepsilon$ %) of $Bi_{1-x}Ca_xFe_{1-y}Ti_yO_{3-\delta}$ nanoparticles with respect to average crystallite size '$d$'. Star symbols in (b) show the microstrain in pure BFO nanoparticles of size ranging from 5 to 65 nm for comparison. (c) Deconvoluted two-phonon mode in selected samples showing the shift in peaks with decreasing size '$d$'. (d) Position of the first peak at ~ 940 cm$^{-1}$ in the two-phonon band plotted as a function of '$d$'.

**Fig. 2 (Color online):** Schematic illustration of different microstrain sources in bismuth ferrite system depicted by the typical cross-sectional atomic arrangement in $(100)_{pc}/(012)_h$ plane. (a) represents the strain-free lattice (b) shows the surface-induced microstrain in nanoparticles and (c) core strain due to the oxygen vacancies in Ca doped $BiFeO_3$ lattice.

**Fig. 3 (Color online):** Differential scanning calorimetry (DSC) data of (a) Ca doped ($y$ =0) and (b) Ca-Ti co-doped ($x = y$) samples. Dotted lines are just a guide to the eye indicating the shift in crystallization peaks.

**Fig. 4:** Bright field images of (a) spherical pure BFO ($d \approx 65$ nm) nanoparticles. Strain field contrast in (b) BCFO-650 ($x = 0.05$, $d \approx 57$ nm) and (c) BCFTO-650 ($x = y = 0.05$, $d \approx 32$ nm). Despite annealed under identical conditions, the size difference in the doped samples with respect to the pure BFO can be seen.

**Fig. 5:** Representative HRTEM strain contrast and strained lattice (filtered FFT) images of BFO (a and b), BCFO (c and d) and BCFTO (e and f). The squares on the HRTEM images (left) represent the selected area for obtaining filtered FFT images. Arrows on filtered FFT images (right) show the strained regions.

**Fig. 6 (Color online):** Room temperature M-H plots of BCFO samples ((a) to (c)) and BCFTO samples ((d) to (f)) for BCFO and BCFTO samples annealed at 550 °C, 650 °C and 800 °C respectively.

**Fig. 7 (Color online):** Ferromagnetic magnetization ($M_f$) of (a) BCFO and (b) BCFTO as a function of doping concentration along with undoped values ($x = y = 0$) at different annealing

temperatures. The open symbols show room temperature (300 K) values and the filled symbols show the corresponding low temperature (20 K) data for selected samples. The dotted lines are just guide to the eye.

**Fig. 8 (Color online):** Zero-field cooled (black open squares) and field cooled curves (red open circles) of (a) pure BFO ($x = y = 0$); (b) $Bi_{1-x}Ca_xFeO_{3-\delta-x/2}$ ($x = 0.05$, $y = 0$); (c) $Bi_{1-x}Ca_xFe_{1-y}Ti_yO_{3-\delta}$ ($x = y = 0.05$); (d) $Bi_{1-x}Ca_xFeO_{3-\delta-x/2}$ ($x = 0.1$, $y = 0$) and (e) $Bi_{1-x}Ca_xFe_{1-y}Ti_yO_{3-\delta}$ ($x = y = 0.1$) measured with an applied field of 500 Oe. The upturn observed in the high temperature annealed samples (BFO-800 in panel (a)) is fitted (blue solid line) with the equation $M(T) = HC/T + M_S(1-BT^{3/2})$. The first term in this equation is due to Curie susceptibility with C as the Curie term and the second term is the spin-wave term with B as the spin-wave stiffness and $M_S$ is the saturation magnetization of the ferromagnetic component (adapted from Panguluri *et al.* [54]). The curve fits well when ~ 2.5 % of Fe ions are considered to be in the paramagnetic state. We choose to fit BFO-800 data since it has a noticeable increase in magnetization is observed at low temperature. In other samples we expect much smaller fraction (< 2 %) of paramagnetic contribution to the magnetization. While this contribution from paramagnetic Fe ions would exist in all the nanoparticles, large magnetization arising from the size-effect would mask this small increase in magnetization from being seen in M *vs* T plots.

**Fig. 9 (Color online):** Schematic representation of spin structure in pure and doped $BiFeO_3$ nanoparticles: (i) spin cycloid in the core with spin disorder on the surface for pure BFO, (ii) breakdown of spin cycloid due to oxygen vacancy, and (iii) breakdown of spin cycloid due to $Ti^{4+}$ substitution along with enhanced surface disorder. The reduction of the particle size from left to right is a representation of experimentally observed data. An increase in surface spin disorder with reduction of size is shown. The mechanism of formation as well as breakdown of spin cycloids are illustrated respectively in the figures below. Purple arrows represent the net magnetic moment per unit cell and open squares the oxygen vacancies. With oxygen vacancies, the superexchange interaction ceases to exist and individual Fe spins orient randomly breaking the spin cycloid. $Ti^{4+}$ ($d^0$) being non-magnetic, the net magnetic moment in a unit cell is the same as that of a single $Fe^{3+}$. These uncompensated magnetic moments due to Ti substitution can also be randomly oriented within the core breaking the spin cycloid and enhancing the magnetization.

**Fig. 10 (Color online):** (a) Rapid decrease in ferromagnetic magnetization observed in $Bi_{1-x}Ca_xFe_{1-y}Ti_yO_{3-\delta}$ nanoparticles (open symbols) as a function of average particle size $d$ nm. Inset is a magnified view of the plot in (a).Solid curve represents the fit using the empirical formula $M \propto (1/d)\varepsilon^\beta$ ($\beta \approx 1.6$). Orange curve (with open square) is the macroscopic magnetization ($M_{(theory)}$) as a function of particle size for particles exhibiting lognormal distribution obtained from equation (iii).(b) Linear variation between the logarithmic saturation magnetization and particle size. Inset in (b) shows the $M_f$ vs. $1/d$ plot for the experimental and the theoretical data. $M_{(theory)}$ shows near linear $1/d$ dependence for smaller particle sizes (< 50 nm). The dashed red line on the experimental data showing a strong deviation from the $1/d$ linear dependence is just a guide to the eye. The + symbol corresponds to the data set for a random combination of temperature, annealing time and doping concentration to achieve finer size control.

**Fig. 11 (Color online):** The macroscopic magnetization ($M_{theory}$) as a function of particle size for particles exhibiting lognormal distribution with various standard deviations (S). The magnetization is obtained from equation (1) in the manuscript.

**Fig. 12 (Color online):** Morphology of (a) as-prepared pure BFO (b) air-annealed BFO (c) and (e) air-annealed samples of BCFO for $x = 0.05$ and 0.1, respectively. (d) and (f) air-annealed samples of BCFTO for $x = y = 0.05$ and 0.1 respectively. The images show nanosized grains in doped samples.

**Fig. 13 (Color online):** (a) Room temperature ferroelectric hysteresis curves P *vs*. E for the SPS pellets. Inset shows the hysteresis loops of BCFTO samples. (b) Leakage current density J *vs*. E characteristics fitted for linear behavior. The slopes derived are labeled respectively.

**Fig. 14 (Color online):** (a) Low temperature (10 K) M *vs*. H curves of SPS pellets. (b) The magnetodielectric response at 10 K (MD %) plotted as a function of applied magnetic field. The data corresponding to BCFO10 and BC10FT10 samples are multiplied by constant factors for easy comparison.

**Fig. 15 (Color online):** The magnetodielectric response of SPS pellets plotted against the square of the magnetization. The curves take a gradual increase and show linearity for applied field above 5000 Oe.

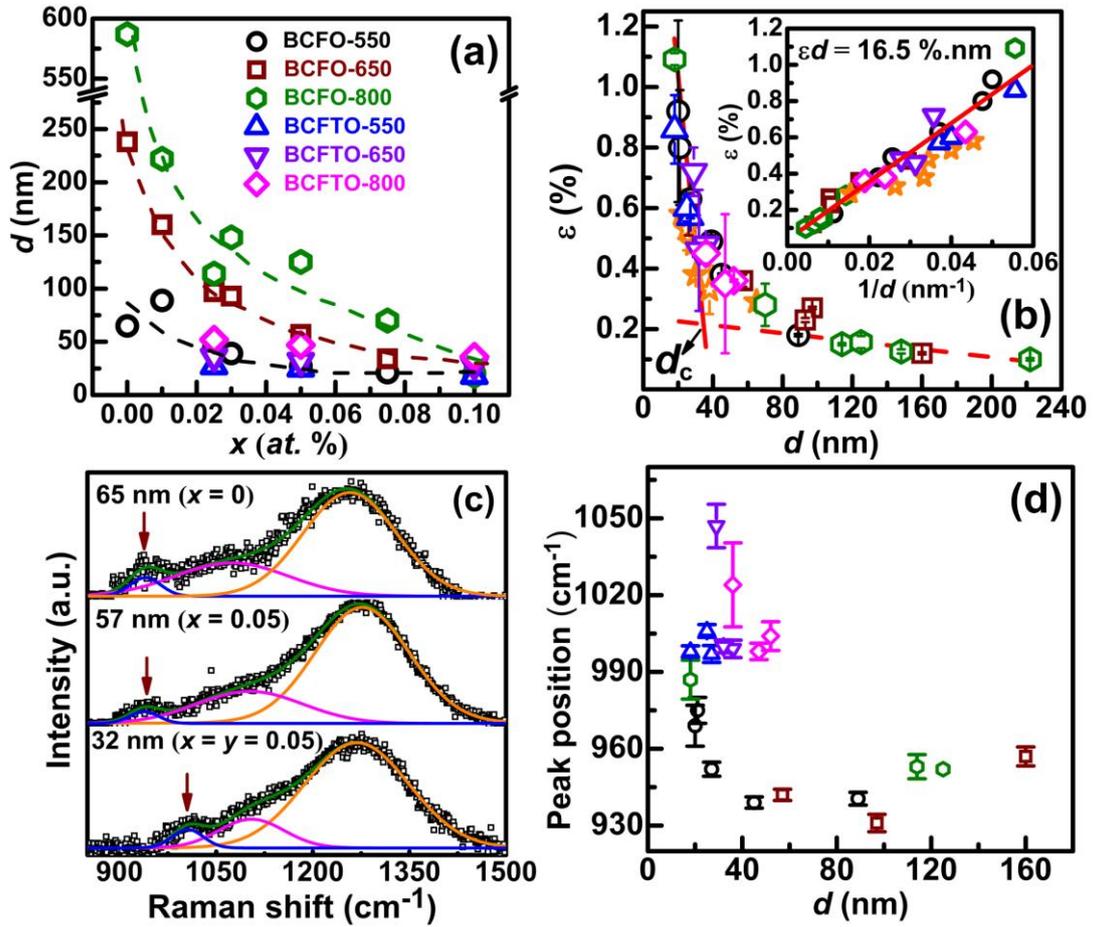

**Fig. 2**

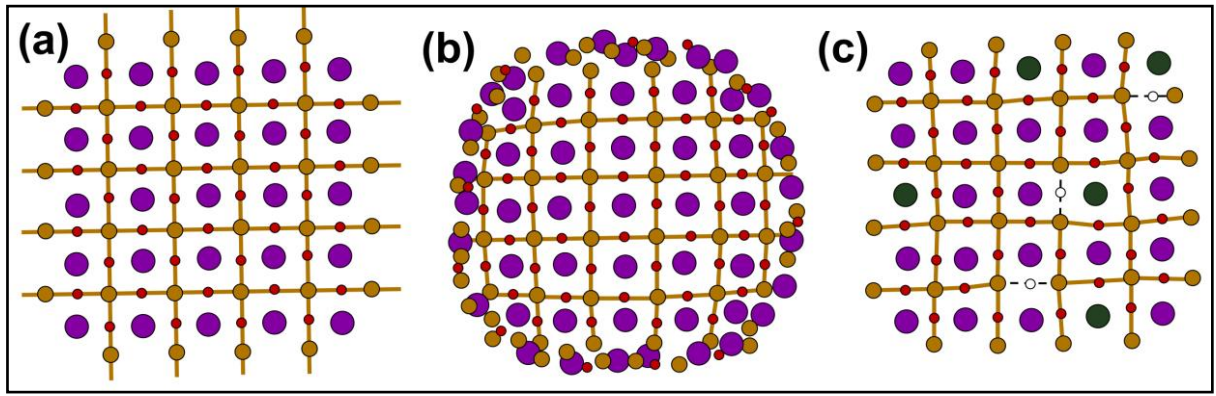

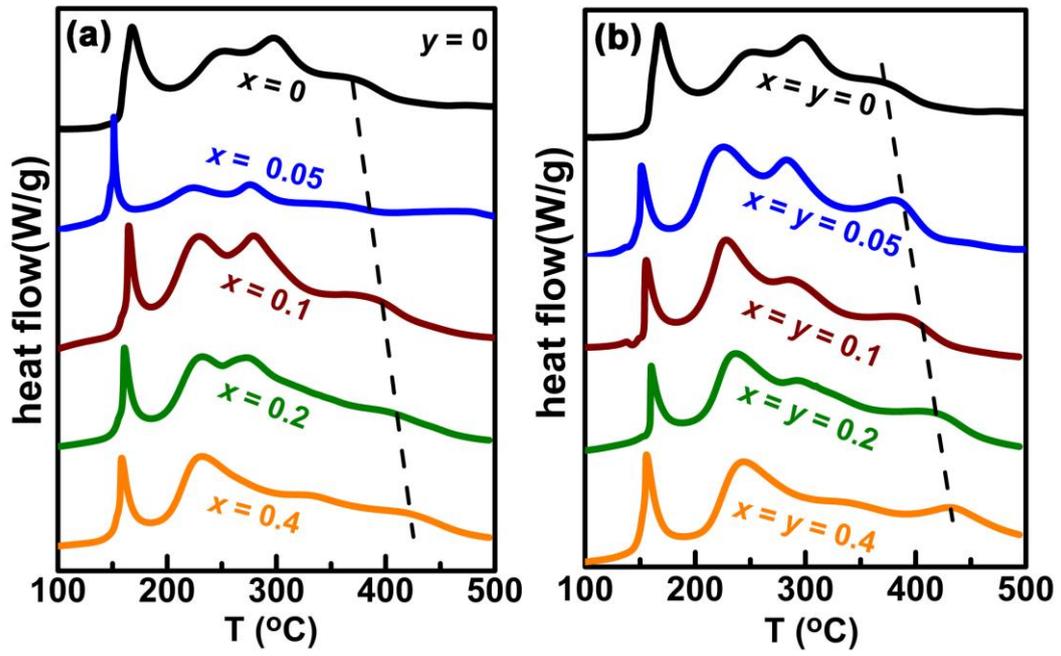

**Fig. 4**

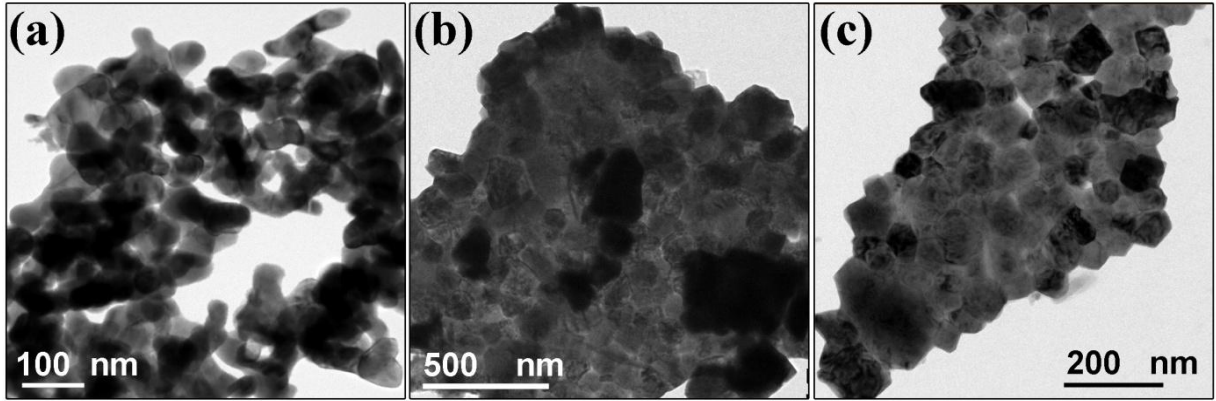

**Fig. 5**

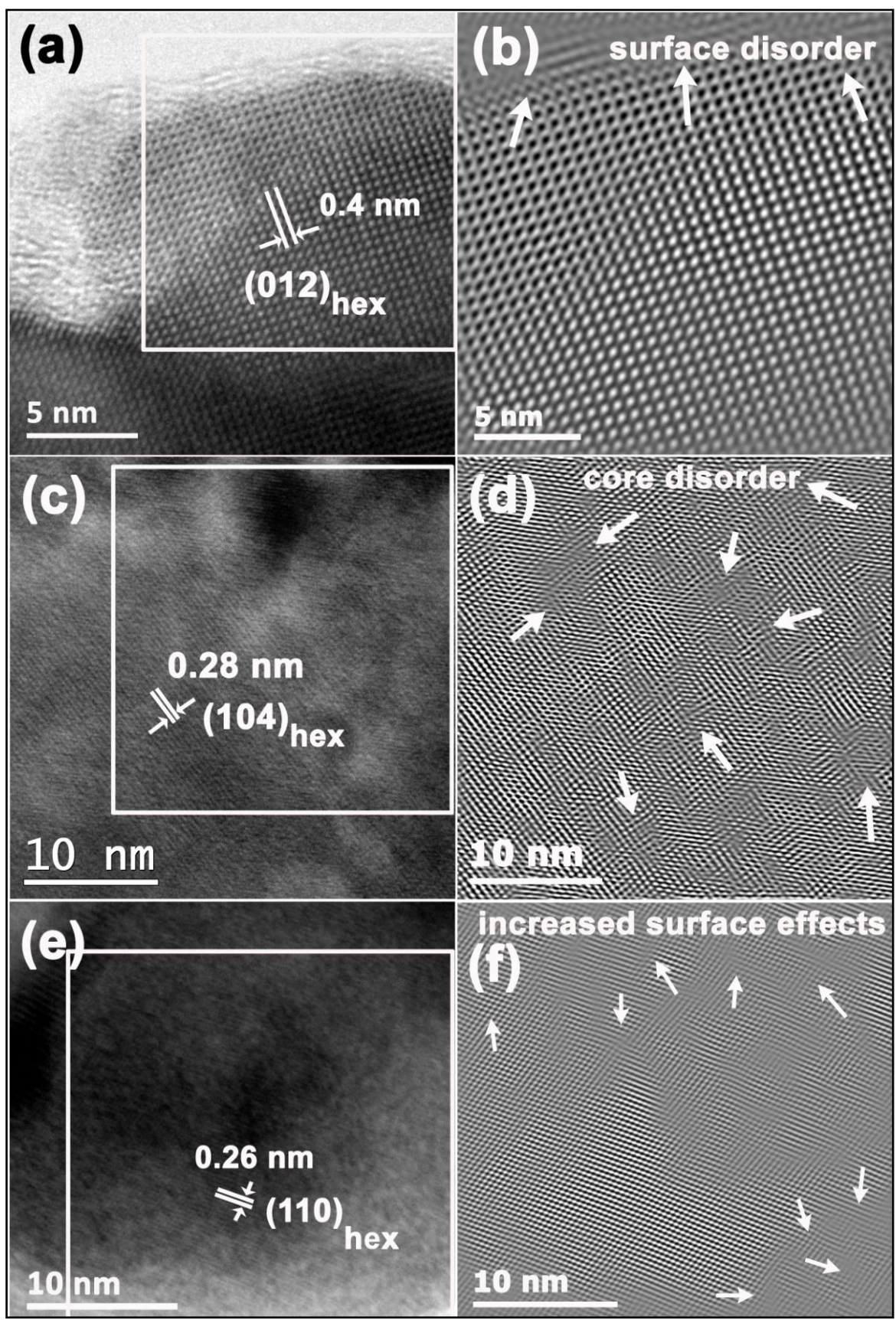

**Fig. 6**

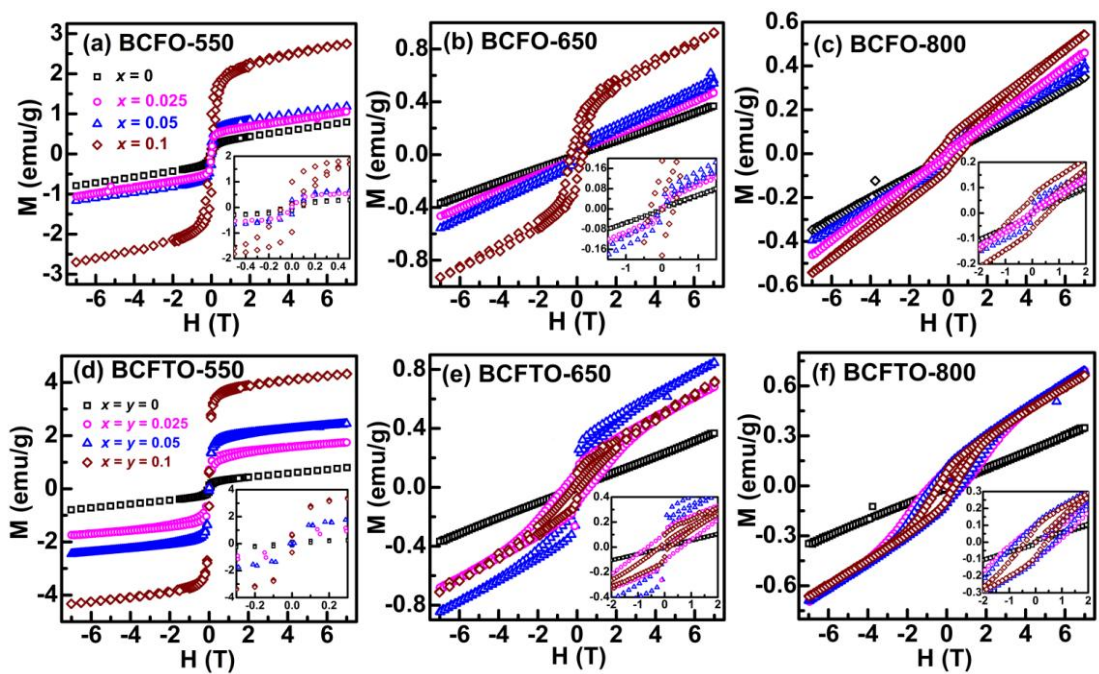

**Fig. 7**

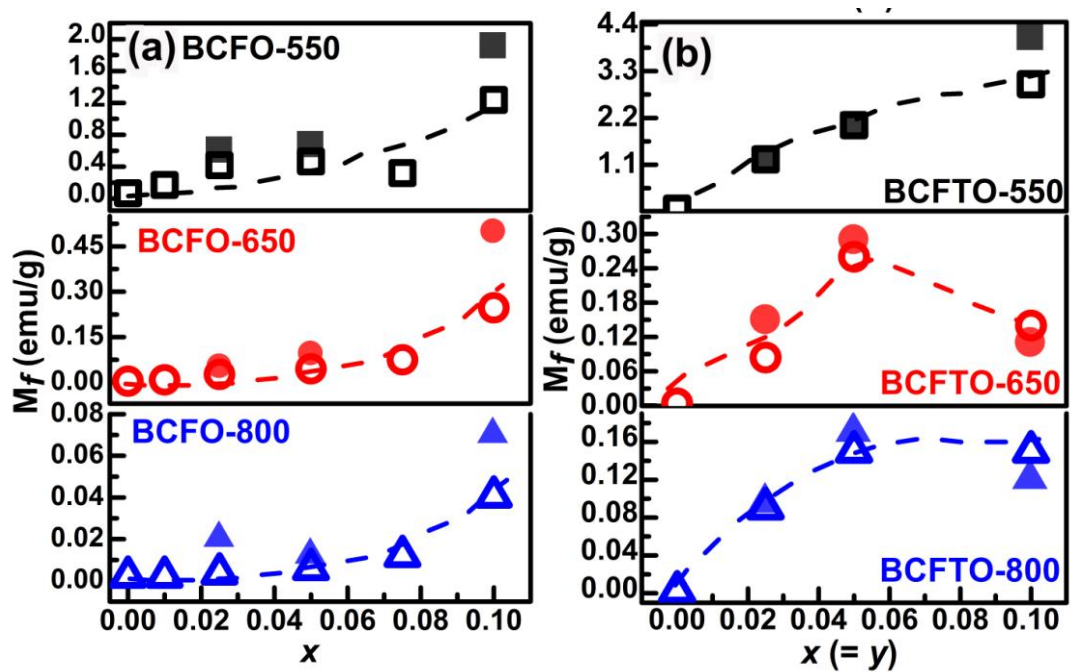

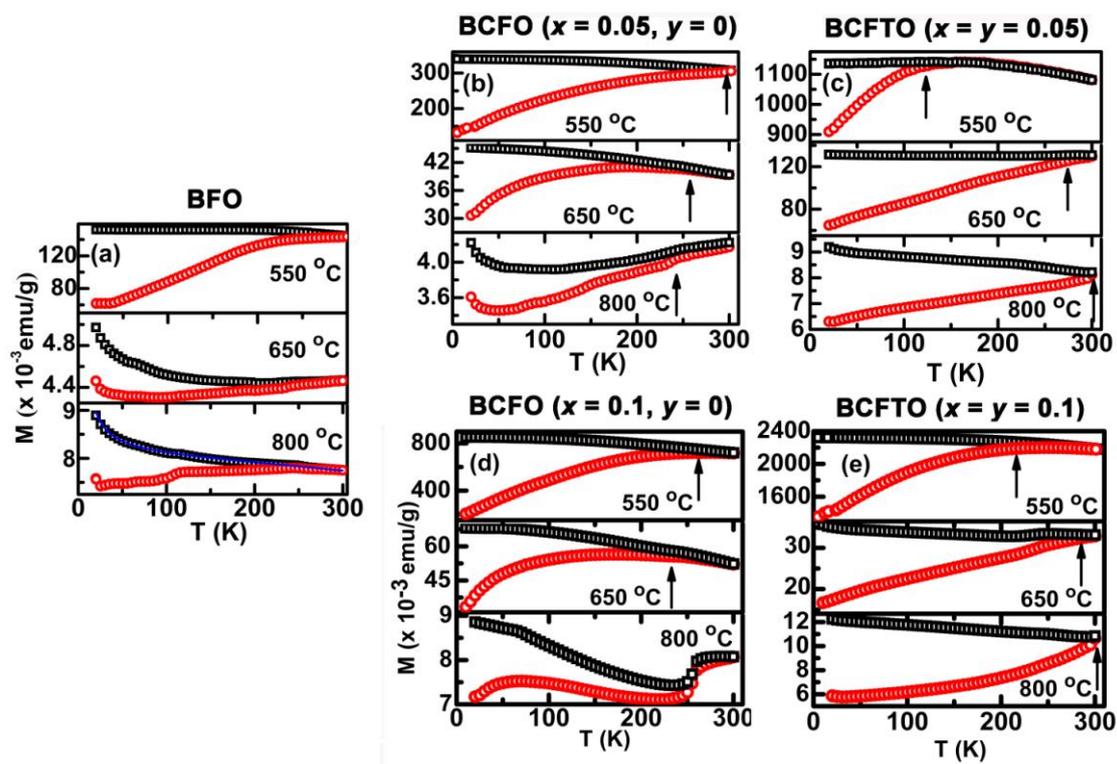

**Fig. 9**

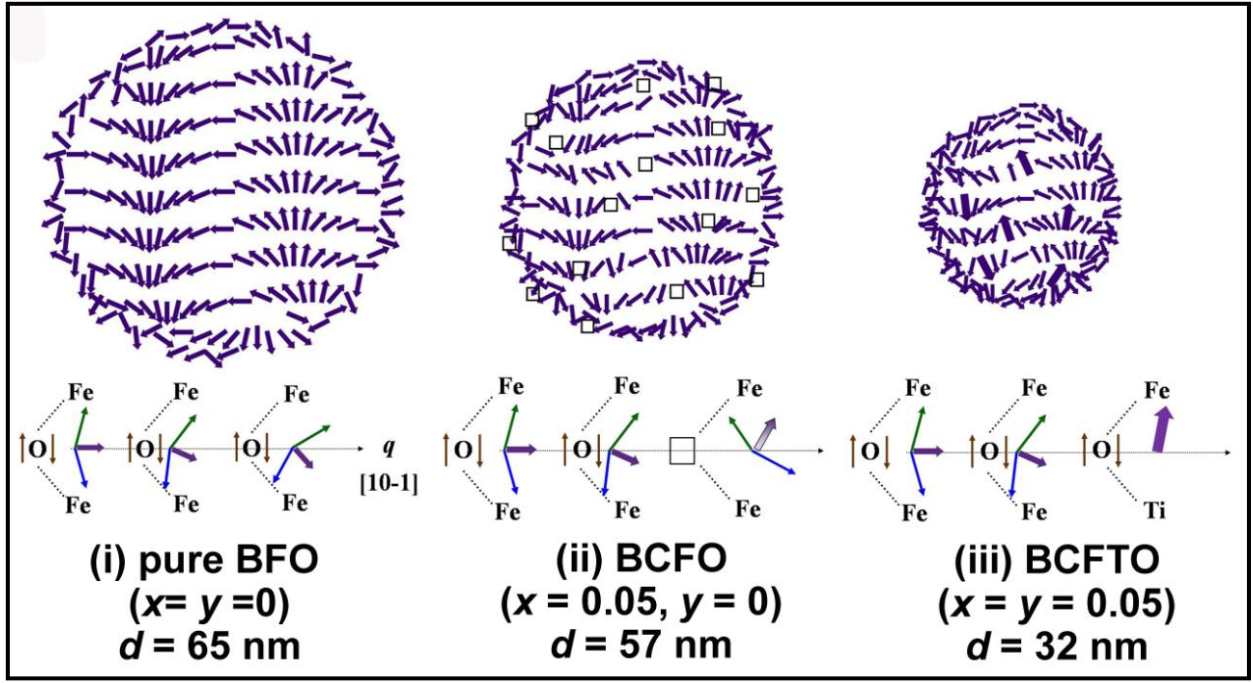

**Fig. 10**

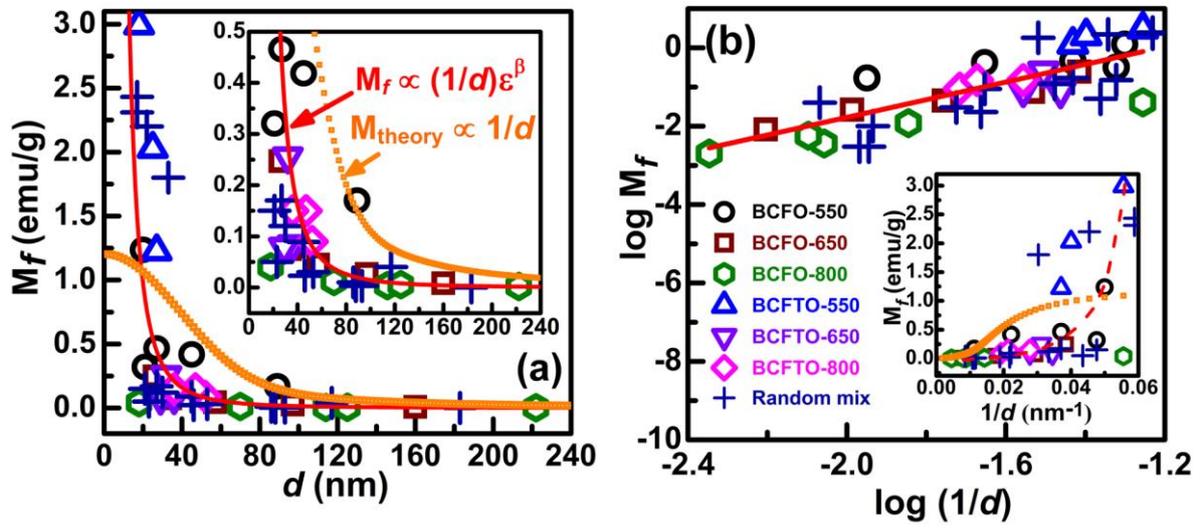

**Fig. 11**

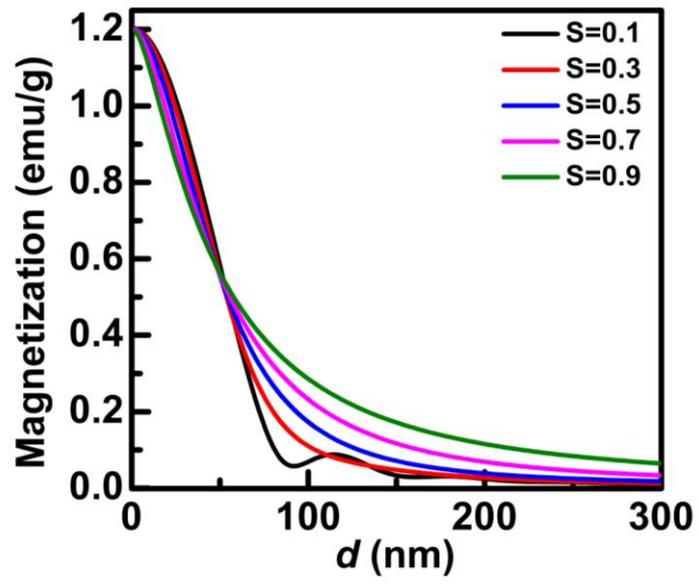

**Fig. 12**

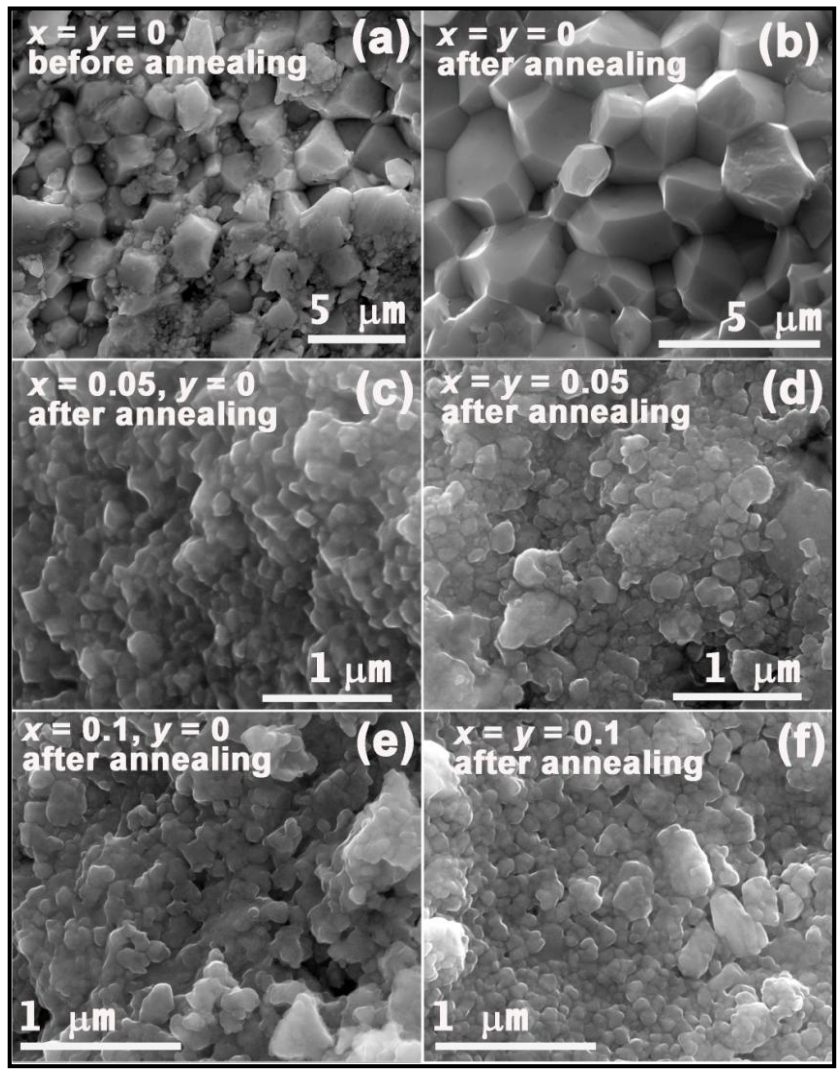

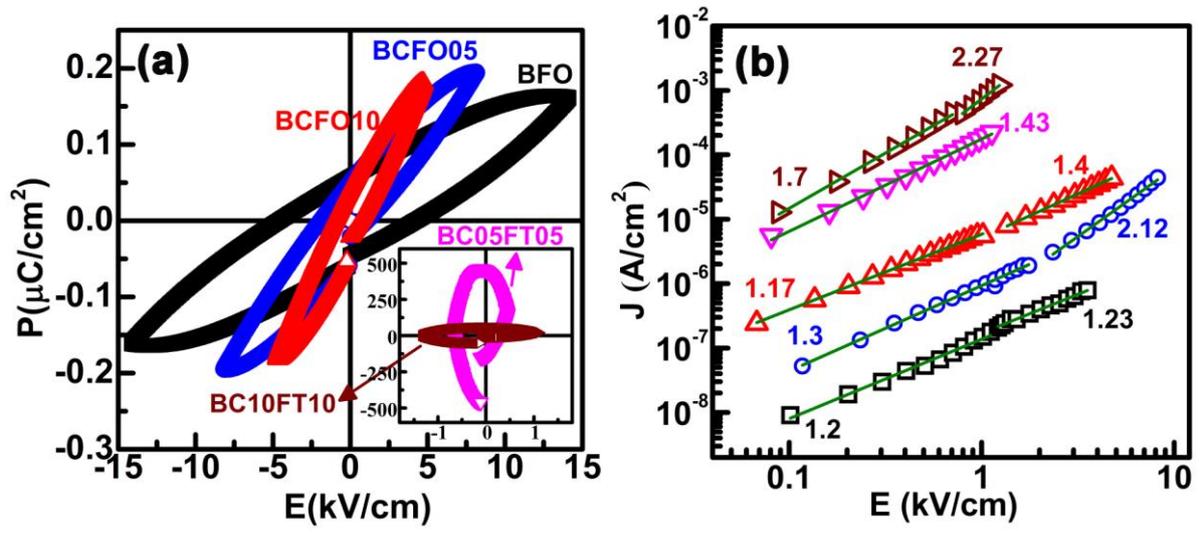

**Fig. 14**

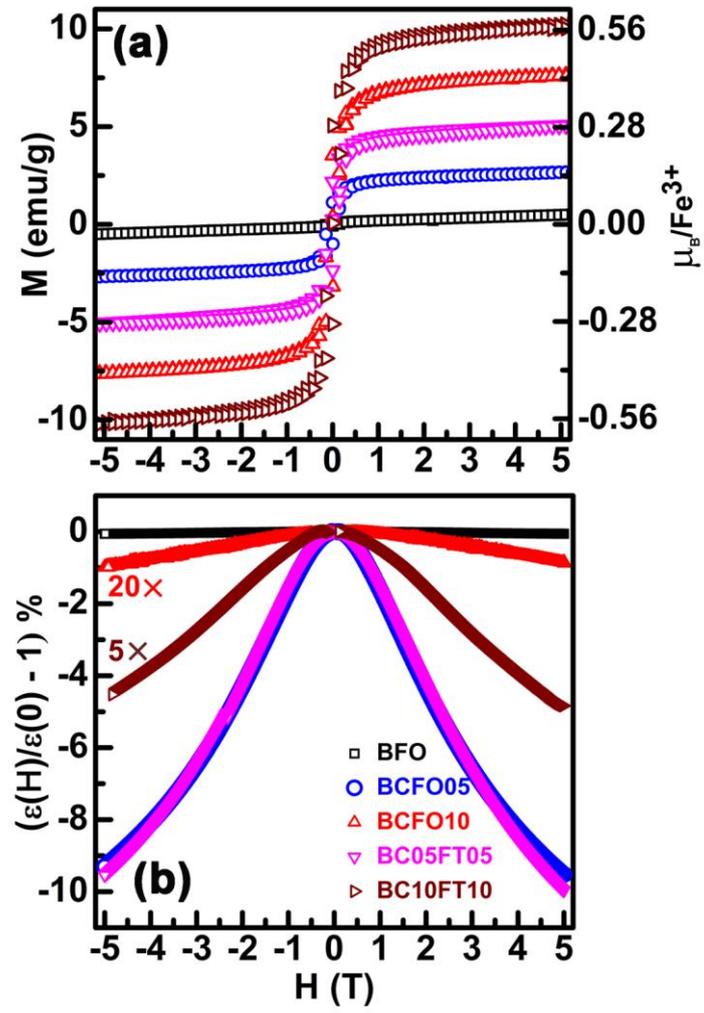

**Fig. 15**

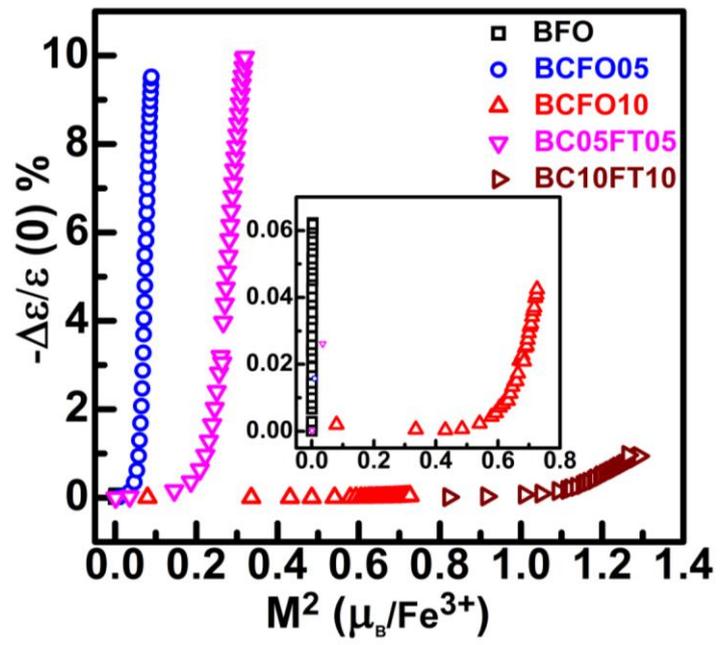